# Nonequilibrium and effect of gas mixtures in an atmospheric microplasma

Davide Mariotti[a)]
*Department of Microelectronic Engineering, Rochester Institute of Technology, Rochester, New York 14623, USA and National Institute of Industrial Science and Technology (AIST), Tsukuba, Ibaraki 305-8568, Japan*

The gas and effective electron temperatures have been estimated for atmospheric microplasma by means of optical emission spectroscopy. The results have shown that the microplasma exhibits nonequilibrium and, as its size is reduced, the two temperatures depart from each other, enhancing the nonequilibrium characteristic. The effect of methane and oxygen concentrations has also been studied, showing that gas mixtures have an important effect on the microplasma state. © *2008 American Institute of Physics*. [DOI: 10.1063/1.2912039]

A quick search in the literature can easily reveal that microplasmas are increasingly grabbing the attention of the scientific world. The search for new technologies and new possibilities is certainly motivating this interest, and the wide range of potential applications are fueling microplasma research. For instance, in material processing, microplasmas can offer unique environments with unprecedented plasma chemistry.[1] Microplasma processing can offer opportunities to synthesize new materials and possibly provide solutions for the future of nanofabrication.[2,3]

Independent of the size of the plasma, the interest in nonequilibrium atmospheric plasmas is justified by the low cost of implementation as well as the possibility of enhancing process rates due to the higher gas density. Research in this area has observed a dramatic increase, and related large-scale nonequilibrium plasmas are commonly referred to as atmospheric plasma glow discharges (APGD). In large-scale low-pressure plasmas, the electron temperature is typically much higher than the gas temperature, exhibiting nonequilibrium. Generally, as the pressure is increased, the two temperatures tend to come closer to each other (as shown by solid lines in Fig. 1) until nonequilibrium disappears and thermodynamic equilibrium is established with typical features of thermal plasmas (Fig. 1).[4] There are conditions for which thermalization does not occur as the pressure is increased. It is speculated that one of this cases is represented by microplasmas where increasing pressure does not imply the loss of nonequilibrium, and at atmospheric pressure, reducing the size of the plasma increases the electron temperature and decreases the gas temperature (as shown by red and blue broken lines in Fig. 1).[4]

Unfortunately, there is still little understanding of the dominant processes in microplasmas and the diagnostic techniques are also limited.[5] Microplasma nonequilibrium for instance, is qualitatively evident but quantitative experimental evidences are still lacking.

In this letter, we will report on the experimental evidence of nonequilibrium in atmospheric microplasma and we will show that nonequilibrium is enhanced as the overall size of the microplasma is reduced. In particular, the gas and effective[6,7] electron temperatures have been estimated on the basis of optical emission spectra. In addition, with the aim to contribute to a better understanding of microplasma processes, the effect of a few gas mixtures on both temperatures will also be analyzed. We have developed, studied, and tested several microplasma configurations with the aim of applying microplasmas for the formation of nanostructures.[1,2,8,9] One of the system configurations used to produce the microplasma is shown in Fig. 2. An ultra high frequency (UHF) (450 MHz) power supply is connected, through a matching network, to the bottom glass epoxy copper plated electrode about $2 \times 2$ mm$^2$ large, which acts as a patch antenna. The substrate is placed on top of the copper electrode for deposition and/or material growth. During experiments, the gas flow is initiated first with the desired composition and rate. The power from the 450 MHz power supply is then set to the specified value. A high voltage pulse of about 5 kV is used to ignite the plasma and is applied to the metallic capillary. The microplasma is then sustained between the capillary exit and the substrate by the UHF power.

The gas temperature was estimated by comparing experimental emission of the $N_2$ rotational band ($C\ ^3\Pi_u$, $\nu_n = 0 \rightarrow B\ ^3\Pi_g$, $\nu_m = 1$) with the calculated spectra. Although pure argon is injected in the capillary, nitrogen lines can be easily observed because the microplasma is generated in ambient air and, therefore, nitrogen easily mixes at the capillary exit. The method has been described in detail and validated elsewhere[10,11] and relies on the assumption that rotational relaxation is much faster than vibrational and electronic transitions. For the electron temperature, a technique described elsewhere[6,7] and based on a collisional-radiative model has been used. This technique has been already applied to micro-

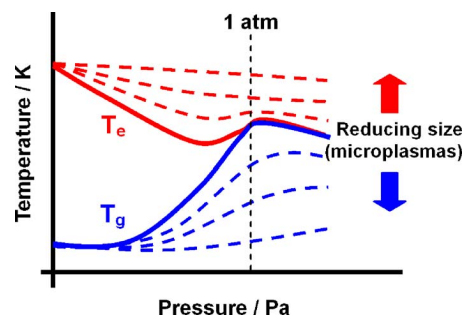

FIG. 1. (Color online) Diagram showing how nonequilibrium is lost when the pressure is increased; nonequilibrium is again reacquired by reducing the size of the plasma (Ref. 2).

---

[a)]Electronic address: davide.mariotti@rit.edu and dxmemc@rit.edu.

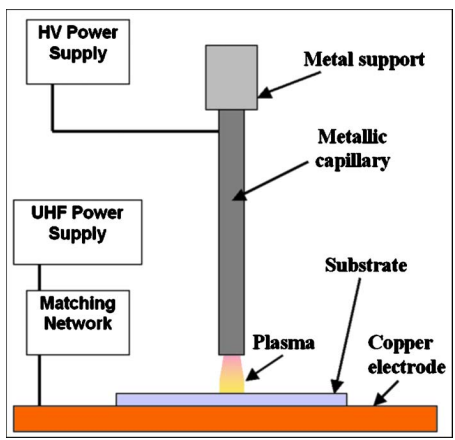

FIG. 2. (Color online) System configuration of the microplasma.

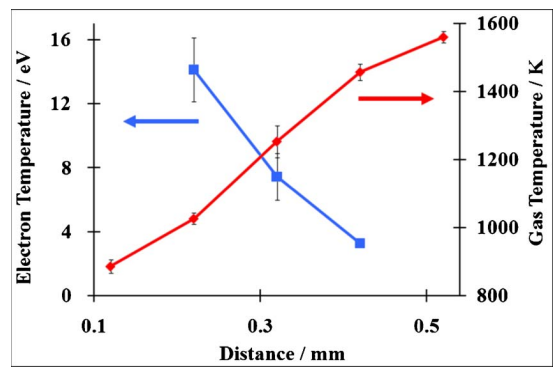

FIG. 3. (Color online) Effective electron and gas temperatures of an atmospheric microplasma vs plasma height (Ar at 20 SCCM; 450 MHz power supply at 7 W).

plasmas and provides the value for an effective electron temperature.[12]

By reducing the distance between the capillary exit and the substrate, the microplasma was reduced in size and, by increasing the distance, the overall microplasma volume was enlarged. For this experiment, a Ni capillary (1 mm external diameter and 0.8 mm internal diameter) was used. The gas was argon at 20 SCCM (SCCM denotes cubic centimeter per minute at STP), the substrate was made out of a Si wafer (0.625 mm thick), and the power was fixed at 7 W. The results are reported in Fig. 3, where electron and gas temperatures are shown for different distances between the capillary exit and the substrate. The error bars represent the measurement standard deviations. Results in Fig. 3 are the experimental evidence of nonequilibrium for this microplasma, whereby electron temperature is much higher than the gas temperature. This is to say that in substance, this microplasma is, in fact, an APGD. Moreover, it can be observed that as the distance is reduced and, consequently, the size of the microplasma gets smaller, the two temperatures depart from each other, enhancing nonequilibrium.

The results can be interpreted in terms of the changes that occur in the production of electrons. As the size of the plasma is reduced (the surface area to volume ratio increases), electron production at the surfaces is much more efficient and the required degree of volume ionization to sustain the plasma diminishes. Ions represent the only energy channel for increasing gas temperature, as the much lighter electrons are unable to transfer energy to heavy molecules and atoms. At these conditions, the ion density is necessarily reduced with consequent decrease of the gas temperature. On the other hand, the increase of the electron temperature by decreasing the plasma size can be justified by a better power coupling from the EM fields to the electrons due to the closer proximity of the emitting capillary surfaces to the copper antenna, which also enhances the energy of emitted secondary electrons.[5] It has to be noted that the variations in the electron energy are described here in terms of an effective electron temperature. Likely, the distribution of electron energy is not Maxwellian so that a shift of electrons toward higher energies may occur through phenomena such as high energy secondary electrons[5] and run-away electrons that are increasingly subjected to forward scattering.

We will now show the results related to the effect of gas mixtures on the microplasma state, as the use of gas mixtures is very important for application in material processing. The same temperature measurement methods used above have been applied here to observe the effect of methane in the gas and electron temperatures (Fig. 4). For these results, the capillary-substrate distance was 0.3 mm and the power applied was 3 W.

The overall effect of methane is to deplete groups of electron at higher energy due to excitation and dissociation processes so that the energy coupled to the electrons is generally transferred more to the formation of radicals obtained from $CH_4$ dissociation. In this sense, the electron energy is, therefore, reduced as methane concentration is increased. For the same reason, ionization is affected and a lower ion production also reduces the gas temperature.

Somewhat different is the behavior observed when oxygen is introduced. Microplasma conditions are the same as above except that this time an argon-oxygen mixture at 20 SCCM is used instead of argon-methane.

The electron temperature is observed to decrease with $O_2$ added (Fig. 5). Similar to methane, the introduction of oxygen in the gas mixture certainly contributes to energy losses for electrons; the number of excitation, ionization, and dissociation processes is quite large, and many of these processes that involve atomic and/or molecular oxygen require relatively low activation energies (4.5–14.7 eV).[13] The decreasing electron temperature can be interpreted in the same way as it was done for the case of argon-methane mixtures. Due to the electron-induced processes with oxygen, less electron energy is available for ionization and generation of argon ions. It follows that argon ions are likely to have re-duced densities when oxygen is introduced and cannot be considered responsible for an increase in gas temperature in this case. Nevertheless, electron collisions with oxygen efficiently promote a variety of positive and negative ions ($O^+$, $O^{2+}$, $O^{4+}$, $O^{3-}$, $O^{2-}$, and $O^-$).[13] These oxygen-related ions can all effectively contribute to increasing gas temperature. In particular, dissociative electron attachment leading to $O^-$ $+O$ is very effective and is activated at energy as low as 5 eV;[13] this dissociative process can reduce electron energy, decrease electron density, and contribute to gas heating all at the same time. The characteristic properties of oxygen and its electronegativity seem to be responsible for transferring energy from electrons to an overall increased gas temperature.

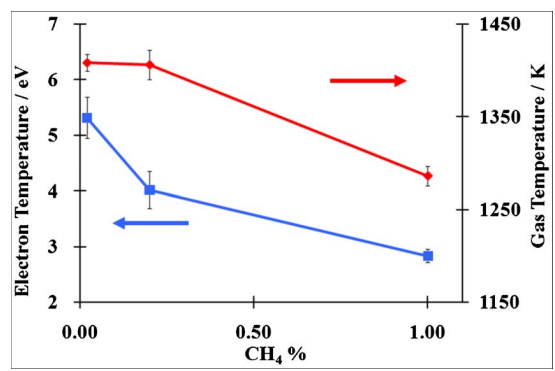

FIG. 4. (Color online) Changes of effective electron and gas temperatures varying methane concentration in Ar background (total gas flow at 20 SCCM; 450 MHz power supply at 3 W; the capillary was 0.3 mm distant from the substrate).

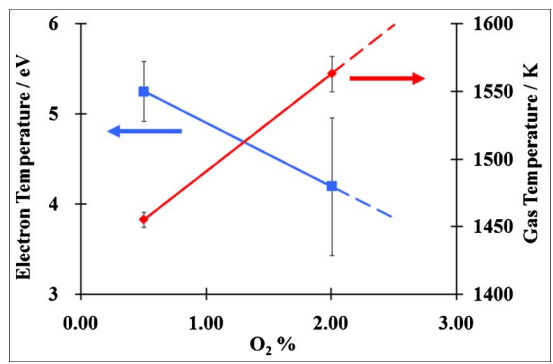

FIG. 5. (Color online) Changes of effective electron and gas temperatures varying oxygen concentration in Ar background (total gas flow at 20 SCCM; 450 MHz power supply at 3 W; the capillary was 0.3 mm distant from the substrate).

The results reported show that atmospheric microplasmas remain APGDs for different important gas mixtures. Moreover, the results underline the importance of gas mixture in determining the microplasma state.

This letter has reported on the experimental evidence of nonequilibrium atmospheric pressure microplasmas and has shown that the degree of nonequilibrium depends on the microplasma size. It has also underlined that, in fact, microplasmas can be classified as APGDs and they preserve nonequilibrium for a range of gas mixtures. All these represent useful attributes for material processing, offering low-cost and unique processing environments that can be used to meet the future requirements of nanofabrication.